\documentclass[sigconf, review=false]{acmart}

\usepackage{booktabs} 

\setcopyright{rightsretained}

\acmDOI{10.475/123_4}

\acmISBN{123-4567-24-567/08/06}

\acmConference[ICVGIP'24]{15th Indian Conference on Computer Vision, Graphics and Image Processing}{December 2024}{Bengaluru, India}
\acmYear{2024}
\copyrightyear{2024}

\acmPrice{15.00}

\begin{document}
\title{Temporal and Spatial Super Resolution with Latent Diffusion Model in Medical MRI images}
\titlenote{Produces the permission block, and
  copyright information}

\author{Vishal Dubey}
\affiliation{%
  \city{Hyderabad}
  \country{India}
  \postcode{000000}
}
\email{vishaldubey0026@gmail.com}

\renewcommand{\shortauthors}{}

\begin{abstract}
Super Resolution (SR) plays a critical role in computer vision, particularly in medical imaging, where hardware and acquisition time constraints often result in low spatial and temporal resolution. While diffusion models have been applied for both spatial and temporal SR, few studies have explored their use for joint spatial and temporal SR, particularly in medical imaging. In this work, we address this gap by proposing to use a Latent Diffusion Model (LDM) combined with a Vector Quantised GAN (VQGAN)-based encoder-decoder architecture for joint super resolution. We frame SR as an image denoising problem, focusing on improving both spatial and temporal resolution in medical images. Using the cardiac MRI dataset from the Data Science Bowl Cardiac Challenge, consisting of 2D cine images with a spatial resolution of 256x256 and 8-14 slices per time-step, we demonstrate the effectiveness of our approach. Our LDM model achieves Peak Signal to Noise Ratio (PSNR) of 30.37, Structural Similarity Index (SSIM) of 0.7580, and Learned Perceptual Image Patch Similarity (LPIPS) of 0.2756, outperforming simple baseline method by 5\% in PSNR, 6.5\% in SSIM, 39\% in LPIPS. Our LDM model generates images with high fidelity and perceptual quality with 15 diffusion steps. These results suggest that LDMs hold promise for advancing super resolution in medical imaging, potentially enhancing diagnostic accuracy and patient outcomes.\footnote{Work done independently, extending M.tech thesis project work}. Code link is also shared.\footnote{https://github.com/vishal-dubey-0026/TemporalSpatialSR\textunderscore{}LDM.git}
\end{abstract}

%
%



\maketitle

\section{Introduction}
Super Resolution (SR) is crucial in medical imaging, where low spatial and temporal resolution can hinder accurate diagnoses due to hardware limitations and acquisition constraints \cite{kim2024data}, \cite{zhou2024spatio}. High-resolution images are especially important in cardiac MRI for assessing heart function and abnormalities. Traditional SR methods, such as interpolation, often fail to capture fine details, particularly in the temporal domain. Recently, diffusion models have shown promise in image restoration tasks by refining images through iterative denoising processes \cite{rombach2022high}, \cite{danier2024ldmvfi}, \cite{lew2024disentangled}, \cite{chen2024learning}, \cite{zhou2024spatio}, \cite{jain2024video}. While diffusion model based approaches have improved SR, most focus on either spatial \cite{chen2024learning}, \cite{zhou2024spatio}, \cite{xing2024simda}, \cite{wang2024exploiting}, \cite{kim2024arbitrary}, \cite{zhou2024upscale}, \cite{gao2023implicit} or temporal SR \cite{danier2024ldmvfi}, \cite{lew2024disentangled}, \cite{jain2024video}, \cite{huang2024motion} independently, with few applying these methods jointly \cite{fu2024global}.
However, their use in medical imaging \cite{li2024rethinking}, particularly for joint spatial and temporal SR, remains limited. In this work, we propose to use a Latent Diffusion Model (LDM)\cite{rombach2022high} combined with a Vector Quantised GAN (VQGAN)-based encoder-decoder architecture\cite{esser2021taming} to address this gap. Our approach formulates SR as an image denoising problem, enabling simultaneous spatial and temporal enhancement of medical images.
We evaluate our method on the cardiac MRI dataset from the Data Science Bowl Cardiac Challenge \cite{MRI} and demonstrate significant improvements in PSNR, SSIM, and LPIPS \cite{moser2401diffusion} compared to baseline models. These results suggest that diffusion models can significantly advance SR in medical imaging, potentially leading to improved diagnostic capabilities.

\section{Experiments}

\begin{figure}[h]

\centering
\includegraphics[width=0.5\columnwidth]{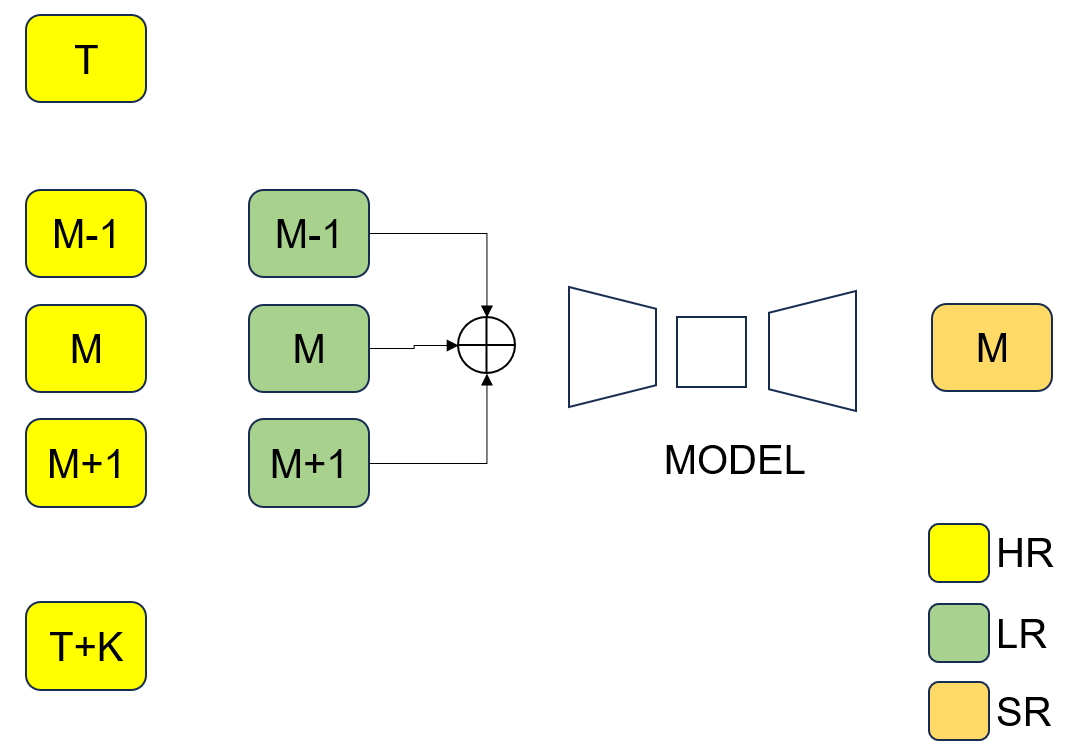}
\caption{temporal and spatial SR framework. Low Resolution (LR) in both temporal and spatial dimension, is obtained by applying OF and spatial degradation to High Resolution (HR) slice. T and M are timesteps, K=8 and $\oplus$ is channel-wise concat}
\Description{Training sample creation from Ground Truth and training model for joint temporal and spatial SR}
\label{fig1}

\end{figure}

\subsection{Training Dataset}
We have used this \textbf{Data Science Bowl Cardiac Challenge Data}\cite{MRI} dataset for our experiments. In this, we have cardiac MRI images in DICOM\footnote{https://en.wikipedia.org/wiki/DICOM} format. These are 2D cine images capturing around 30 volumes throughout the cardiac cycle. The dataset includes MRI scans from about 1,140 patients, with each patient having between 8 and 14 slices per time step. These 2D cine images generally have a spatial resolution of 256x256. Therefore, each patient’s data covers roughly 30 volumes during one complete cardiac cycle. We have used train split and test split for training which contains patient data of 500 subjects and 440 subjects, respectively. \\


\begin{figure}[h]

\begin{center}

\includegraphics[width=1.0\columnwidth]{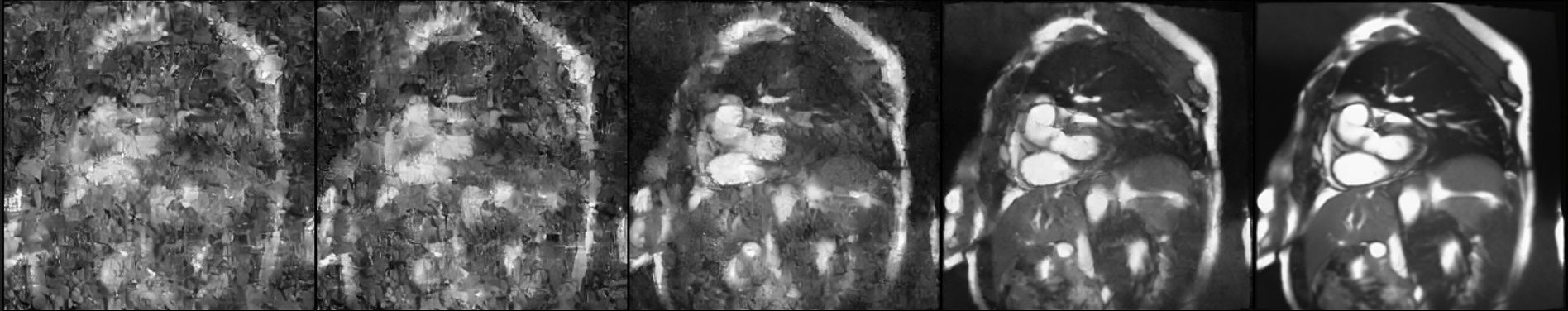}\\
\includegraphics[width=1.0\columnwidth]{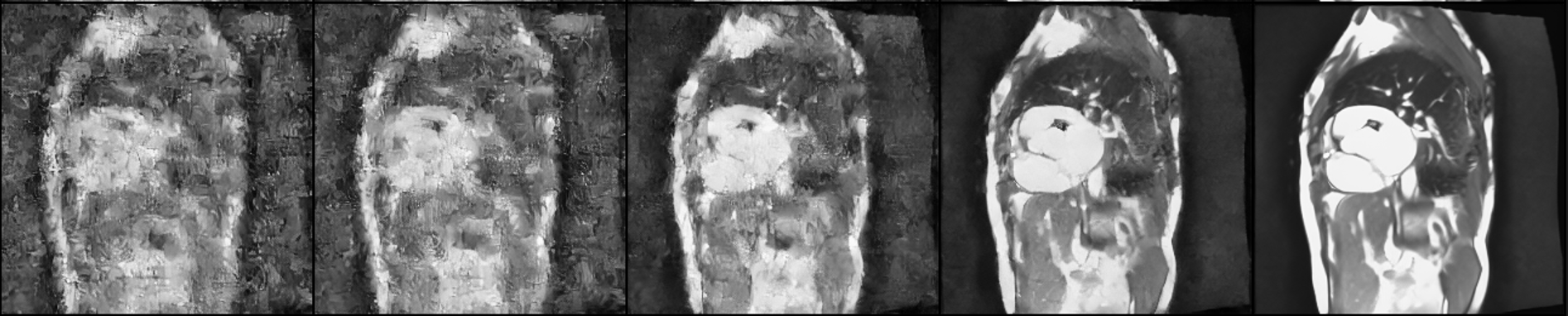}

\caption[width=0.5\columnwidth]{Reverse diffusion(denoising), t=[0,0.25,0.5,0.75,1]}
\label{fig2}
\end{center}
\end{figure}

\begin{figure}[h]

\begin{center}

\includegraphics[width=1.0\columnwidth]{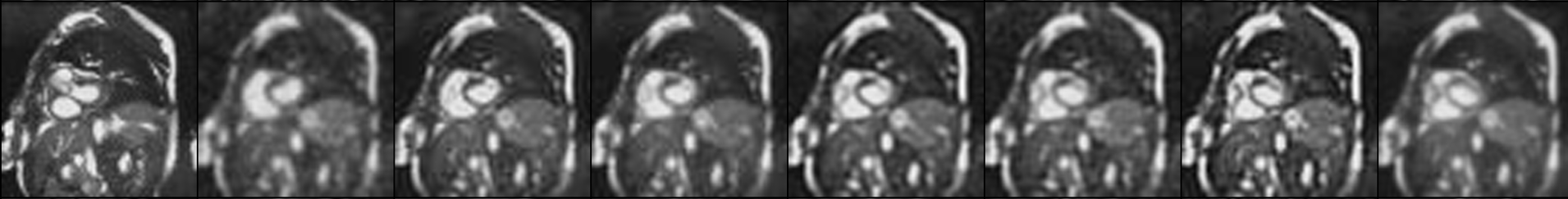}
\includegraphics[width=1.0\columnwidth]{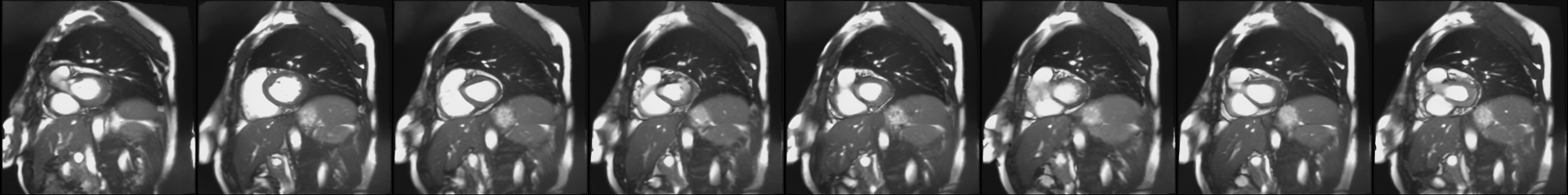}
\includegraphics[width=1.0\columnwidth]{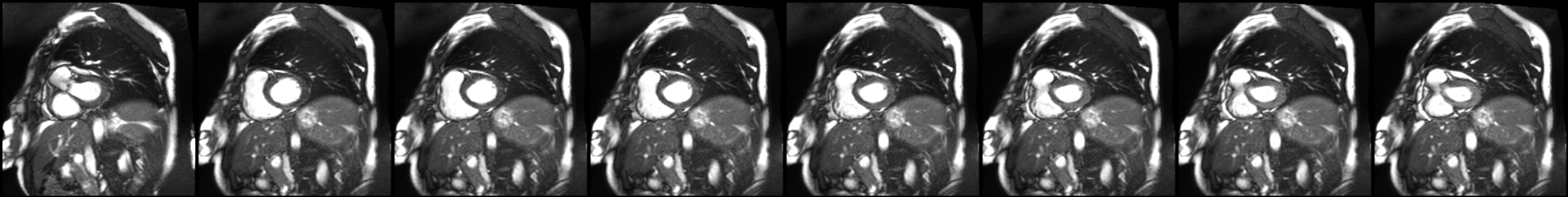}

\caption{Top-to-bottom order(Baseline, our LDM, Ground Truth) and left-to-right order is time}
\label{fig3}
\end{center}

\end{figure}

\begin{figure}[h]

\begin{center}

\includegraphics[width=1.0\columnwidth]{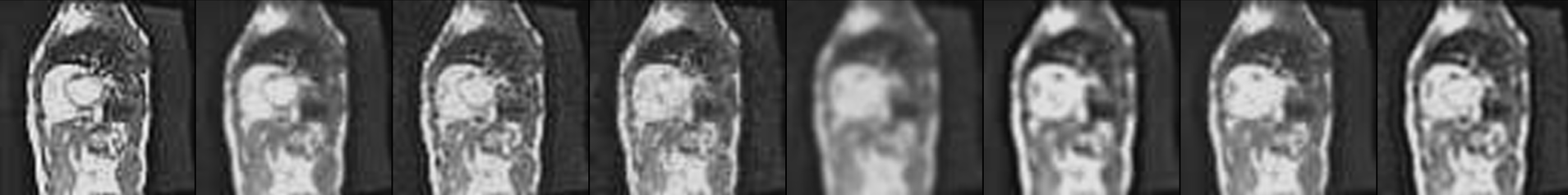}
\includegraphics[width=1.0\columnwidth]{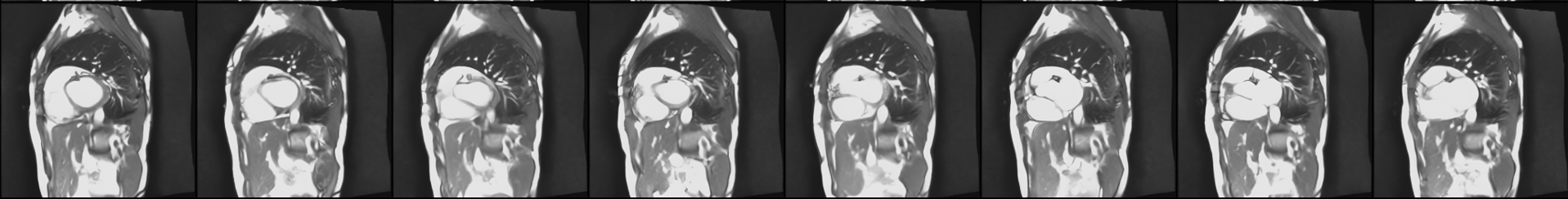}
\includegraphics[width=1.0\columnwidth]{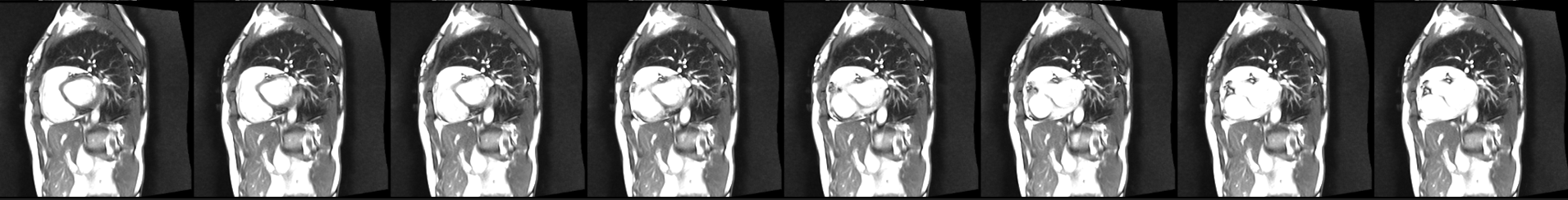}

\caption{Top-to-bottom order(Baseline, our LDM, Ground Truth) and left-to-right order is time}
\label{fig4}
\end{center}

\end{figure}


\subsection{Training Method}
Temporal SR and spatial SR follows frame interpolation framework like \cite{huang2024motion}, \cite{jain2024video}, \cite{danier2024ldmvfi} etc. and image/video SR framework like \cite{zhou2024upscale}, \cite{xing2024simda} etc., respectively. We have formulated joint temporal and spatial SR as an image/frame denoising problem. Spatial SR is done for each slice independently and does not consider the number of slices as depth dimension. Overall framework of temporal and spatial SR is given in Figure \ref{fig1}. We consider a slice for the entire cardiac cycle, with 30 2D cine images as a video-clip with 30 frames. We randomly select a mini-clip of 9 consecutive frames. We take first and ninth frame as end points for frame interpolation. We did frame interpolation using Optical Flow (OF)\cite{farneback2003two} method to generate intermediate low resolution frames in temporal dimension and took random, 3 consecutive interpolated frames. These 3 consecutive frames are spatially degraded/downscaled using degradation pipeline of \cite{realesrgan}, referred to as low resolution input. Ground Truth (GT) is the 3 original consecutive frames(corresponding to low resolution input's timesteps) before frame interpolation. We have resized 2D cine images to 256x256 spatial resolution for training. This kind of framework enables our model to get temporal information of neighboring frames.
We have used architecture given by this paper \cite{yue2024resshift} and loss function of equation-8 of \cite{yue2024resshift} to train the model and have used 15 diffusion steps for training and inference. We have used learning rate of 5e-5 with Adam optimizer, total batch size of 176 with gradient accumulation of 4 steps. We have trained the model for 25000 iterations on single A100 GPU. LDM and VQGAN weights are initialized from weights given by \cite{yue2024resshift}, with VQGAN weights being frozen. We have trained the model for 8x temporal SR and 4x spatial SR.

\begin{table}[htbp]
    \caption{4x spatial, 8x temporal patient200}
    \begin{center}

    \begin{tabular}[scale=1.0]{|c|c|c|c|}
 \hline
\textbf{Models} & \textbf{PSNR}\bf $\uparrow$ & \textbf{SSIM} \bf $\uparrow$ &
\textbf{LPIPS} \bf $\downarrow$\\
 \hline
 Baseline$^{\mathrm{a}}$ & 28.89 & 0.7113 & 0.4536\\
  \hline
 our LDM$^{\mathrm{a}}$ & 30.37 & 0.7580 & 0.2756\\
  \hline
   Baseline$^{\mathrm{b}}$ & 34.30 & 0.9102 & 0.2508\\
  \hline
 our LDM$^{\mathrm{b}}$ & 33.73 & 0.8433 & 0.1785\\
\hline
\multicolumn{3}{p{200pt}}{$^{\mathrm{a}}$ Spatial degradation using ESRGAN pipeline \cite{realesrgan}.}\\
\multicolumn{3}{p{200pt}}{$^{\mathrm{b}}$ Spatial degradation with bicubic downsampling only.}
\end{tabular}
\label{tab1}
\end{center}

\end{table}

\section{Quantitative Evaluation and Results}
temporal and spatial SR is evaluated on validation split of cardiac-MRI dataset \cite{MRI} and it contains patient data of 200 subjects ("patient200"). We have evaluated our LDM for spatial SR of 4x and temporal SR of 8x. We have compared our model against simple \textbf{Baseline} method which is Bicubic-upscaled version of our model input. Table \ref{tab1} reports experiment results. The results in Table \ref{tab1} are of intermediate model checkpoint, with ongoing training. Figure \ref{fig2} shows reverse diffusion during inference and Figure \ref{fig3} \& Figure \ref{fig4} shows SR results.

\section{CONCLUSIONS and Future work}
We can see the effectiveness of diffusion models for joint temporal and spatial SR in medical scans. Diffusion models can do SR of degraded and low quality medical scans, maintaining high fidelity and perceptual quality. Our future work will be to reduce number of denoising steps in diffusion model, using techniques like distillation\cite{wang2024sinsr}, \cite{luo2023latent}. One more future work can be to extend LDM and VQGAN for processing 3D volumes using 3D convolution layers \cite{kim2024data}, analysing their effectiveness.


\bibliographystyle{ACM-Reference-Format}
\bibliography{cite}


\begin{thebibliography}{23}


\ifx \showCODEN    \undefined \def \showCODEN     #1{\unskip}     \fi
\ifx \showDOI      \undefined \def \showDOI       #1{#1}\fi
\ifx \showISBNx    \undefined \def \showISBNx     #1{\unskip}     \fi
\ifx \showISBNxiii \undefined \def \showISBNxiii  #1{\unskip}     \fi
\ifx \showISSN     \undefined \def \showISSN      #1{\unskip}     \fi
\ifx \showLCCN     \undefined \def \showLCCN      #1{\unskip}     \fi
\ifx \shownote     \undefined \def \shownote      #1{#1}          \fi
\ifx \showarticletitle \undefined \def \showarticletitle #1{#1}   \fi
\ifx \showURL      \undefined \def \showURL       {\relax}        \fi
\providecommand\bibfield[2]{#2}
\providecommand\bibinfo[2]{#2}
\providecommand\natexlab[1]{#1}
\providecommand\showeprint[2][]{arXiv:#2}

\bibitem[\protect\citeauthoryear{Chen, Long, Qiu, Yao, Zhou, Luo, and Mei}{Chen et~al\mbox{.}}{2024}]%
        {chen2024learning}
\bibfield{author}{\bibinfo{person}{Zhikai Chen}, \bibinfo{person}{Fuchen Long}, \bibinfo{person}{Zhaofan Qiu}, \bibinfo{person}{Ting Yao}, \bibinfo{person}{Wengang Zhou}, \bibinfo{person}{Jiebo Luo}, {and} \bibinfo{person}{Tao Mei}.} \bibinfo{year}{2024}\natexlab{}.
\newblock \showarticletitle{Learning Spatial Adaptation and Temporal Coherence in Diffusion Models for Video Super-Resolution}. In \bibinfo{booktitle}{\emph{Proceedings of the IEEE/CVF Conference on Computer Vision and Pattern Recognition}}. \bibinfo{pages}{9232--9241}.
\newblock


\bibitem[\protect\citeauthoryear{Danier, Zhang, and Bull}{Danier et~al\mbox{.}}{2024}]%
        {danier2024ldmvfi}
\bibfield{author}{\bibinfo{person}{Duolikun Danier}, \bibinfo{person}{Fan Zhang}, {and} \bibinfo{person}{David Bull}.} \bibinfo{year}{2024}\natexlab{}.
\newblock \showarticletitle{Ldmvfi: Video frame interpolation with latent diffusion models}. In \bibinfo{booktitle}{\emph{Proceedings of the AAAI Conference on Artificial Intelligence}}, Vol.~\bibinfo{volume}{38}. \bibinfo{pages}{1472--1480}.
\newblock


\bibitem[\protect\citeauthoryear{Esser, Rombach, and Ommer}{Esser et~al\mbox{.}}{2021}]%
        {esser2021taming}
\bibfield{author}{\bibinfo{person}{Patrick Esser}, \bibinfo{person}{Robin Rombach}, {and} \bibinfo{person}{Bjorn Ommer}.} \bibinfo{year}{2021}\natexlab{}.
\newblock \showarticletitle{Taming transformers for high-resolution image synthesis}. In \bibinfo{booktitle}{\emph{Proceedings of the IEEE/CVF conference on computer vision and pattern recognition}}. \bibinfo{pages}{12873--12883}.
\newblock


\bibitem[\protect\citeauthoryear{Farneb{\"a}ck}{Farneb{\"a}ck}{2003}]%
        {farneback2003two}
\bibfield{author}{\bibinfo{person}{Gunnar Farneb{\"a}ck}.} \bibinfo{year}{2003}\natexlab{}.
\newblock \showarticletitle{Two-frame motion estimation based on polynomial expansion}. In \bibinfo{booktitle}{\emph{Image Analysis: 13th Scandinavian Conference, SCIA 2003 Halmstad, Sweden, June 29--July 2, 2003 Proceedings 13}}. Springer, \bibinfo{pages}{363--370}.
\newblock


\bibitem[\protect\citeauthoryear{Fu, Yuan, Jiang, Zhang, Shen, and Hamzaoui}{Fu et~al\mbox{.}}{2024}]%
        {fu2024global}
\bibfield{author}{\bibinfo{person}{Congrui Fu}, \bibinfo{person}{Hui Yuan}, \bibinfo{person}{Shiqi Jiang}, \bibinfo{person}{Guanghui Zhang}, \bibinfo{person}{Liquan Shen}, {and} \bibinfo{person}{Raouf Hamzaoui}.} \bibinfo{year}{2024}\natexlab{}.
\newblock \showarticletitle{Global Spatial-Temporal Information-based Residual ConvLSTM for Video Space-Time Super-Resolution}.
\newblock \bibinfo{journal}{\emph{arXiv preprint arXiv:2407.08466}}.
\newblock


\bibitem[\protect\citeauthoryear{Gao, Liu, Zeng, Xu, Li, Luo, Liu, Zhen, and Zhang}{Gao et~al\mbox{.}}{2023}]%
        {gao2023implicit}
\bibfield{author}{\bibinfo{person}{Sicheng Gao}, \bibinfo{person}{Xuhui Liu}, \bibinfo{person}{Bohan Zeng}, \bibinfo{person}{Sheng Xu}, \bibinfo{person}{Yanjing Li}, \bibinfo{person}{Xiaoyan Luo}, \bibinfo{person}{Jianzhuang Liu}, \bibinfo{person}{Xiantong Zhen}, {and} \bibinfo{person}{Baochang Zhang}.} \bibinfo{year}{2023}\natexlab{}.
\newblock \showarticletitle{Implicit diffusion models for continuous super-resolution}. In \bibinfo{booktitle}{\emph{Proceedings of the IEEE/CVF conference on computer vision and pattern recognition}}. \bibinfo{pages}{10021--10030}.
\newblock


\bibitem[\protect\citeauthoryear{Huang, Yu, Yang, Qin, Zheng, Zheng, Zhou, Wang, and Yang}{Huang et~al\mbox{.}}{2024}]%
        {huang2024motion}
\bibfield{author}{\bibinfo{person}{Zhilin Huang}, \bibinfo{person}{Yijie Yu}, \bibinfo{person}{Ling Yang}, \bibinfo{person}{Chujun Qin}, \bibinfo{person}{Bing Zheng}, \bibinfo{person}{Xiawu Zheng}, \bibinfo{person}{Zikun Zhou}, \bibinfo{person}{Yaowei Wang}, {and} \bibinfo{person}{Wenming Yang}.} \bibinfo{year}{2024}\natexlab{}.
\newblock \showarticletitle{Motion-aware Latent Diffusion Models for Video Frame Interpolation}.
\newblock \bibinfo{journal}{\emph{arXiv preprint arXiv:2404.13534}}.
\newblock


\bibitem[\protect\citeauthoryear{Jain, Watson, Tabellion, Poole, Kontkanen, et~al\mbox{.}}{Jain et~al\mbox{.}}{2024}]%
        {jain2024video}
\bibfield{author}{\bibinfo{person}{Siddhant Jain}, \bibinfo{person}{Daniel Watson}, \bibinfo{person}{Eric Tabellion}, \bibinfo{person}{Ben Poole}, \bibinfo{person}{Janne Kontkanen}, {et~al\mbox{.}}} \bibinfo{year}{2024}\natexlab{}.
\newblock \showarticletitle{Video interpolation with diffusion models}. In \bibinfo{booktitle}{\emph{Proceedings of the IEEE/CVF Conference on Computer Vision and Pattern Recognition}}. \bibinfo{pages}{7341--7351}.
\newblock


\bibitem[\protect\citeauthoryear{kaggle.com}{kaggle.com}{2020}]%
        {MRI}
\bibfield{author}{\bibinfo{person}{kaggle.com}.} \bibinfo{year}{2020}\natexlab{}.
\newblock \bibinfo{title}{Data Science Bowl Cardiac Challenge Data}.
\newblock
\newblock
\urldef\tempurl%
\url{https://www.kaggle.com/c/second-annual-data-science-bowl}
\showURL{%
Retrieved August 1, 2024 from \tempurl}


\bibitem[\protect\citeauthoryear{Kim and Kim}{Kim and Kim}{2024}]%
        {kim2024arbitrary}
\bibfield{author}{\bibinfo{person}{Jinseok Kim} {and} \bibinfo{person}{Tae-Kyun Kim}.} \bibinfo{year}{2024}\natexlab{}.
\newblock \showarticletitle{Arbitrary-Scale Image Generation and Upsampling using Latent Diffusion Model and Implicit Neural Decoder}. In \bibinfo{booktitle}{\emph{Proceedings of the IEEE/CVF Conference on Computer Vision and Pattern Recognition}}. \bibinfo{pages}{9202--9211}.
\newblock


\bibitem[\protect\citeauthoryear{Kim, Yoon, Park, Kim, and Yang}{Kim et~al\mbox{.}}{2024}]%
        {kim2024data}
\bibfield{author}{\bibinfo{person}{JungEun Kim}, \bibinfo{person}{Hangyul Yoon}, \bibinfo{person}{Geondo Park}, \bibinfo{person}{Kyungsu Kim}, {and} \bibinfo{person}{Eunho Yang}.} \bibinfo{year}{2024}\natexlab{}.
\newblock \showarticletitle{Data-Efficient Unsupervised Interpolation Without Any Intermediate Frame for 4D Medical Images}. In \bibinfo{booktitle}{\emph{Proceedings of the IEEE/CVF Conference on Computer Vision and Pattern Recognition}}. \bibinfo{pages}{11353--11364}.
\newblock


\bibitem[\protect\citeauthoryear{Lew, Choi, Shin, Jung, and Yoon}{Lew et~al\mbox{.}}{2024}]%
        {lew2024disentangled}
\bibfield{author}{\bibinfo{person}{Jaihyun Lew}, \bibinfo{person}{Jooyoung Choi}, \bibinfo{person}{Chaehun Shin}, \bibinfo{person}{Dahuin Jung}, {and} \bibinfo{person}{Sungroh Yoon}.} \bibinfo{year}{2024}\natexlab{}.
\newblock \showarticletitle{Disentangled Motion Modeling for Video Frame Interpolation}.
\newblock \bibinfo{journal}{\emph{arXiv preprint arXiv:2406.17256}}.
\newblock


\bibitem[\protect\citeauthoryear{Li, Rao, Mo, Zhang, Xing, and Zhao}{Li et~al\mbox{.}}{2024}]%
        {li2024rethinking}
\bibfield{author}{\bibinfo{person}{Guangyuan Li}, \bibinfo{person}{Chen Rao}, \bibinfo{person}{Juncheng Mo}, \bibinfo{person}{Zhanjie Zhang}, \bibinfo{person}{Wei Xing}, {and} \bibinfo{person}{Lei Zhao}.} \bibinfo{year}{2024}\natexlab{}.
\newblock \showarticletitle{Rethinking diffusion model for multi-contrast mri super-resolution}. In \bibinfo{booktitle}{\emph{Proceedings of the IEEE/CVF Conference on Computer Vision and Pattern Recognition}}. \bibinfo{pages}{11365--11374}.
\newblock


\bibitem[\protect\citeauthoryear{Luo, Tan, Huang, Li, and Zhao}{Luo et~al\mbox{.}}{2023}]%
        {luo2023latent}
\bibfield{author}{\bibinfo{person}{Simian Luo}, \bibinfo{person}{Yiqin Tan}, \bibinfo{person}{Longbo Huang}, \bibinfo{person}{Jian Li}, {and} \bibinfo{person}{Hang Zhao}.} \bibinfo{year}{2023}\natexlab{}.
\newblock \showarticletitle{Latent consistency models: Synthesizing high-resolution images with few-step inference}.
\newblock \bibinfo{journal}{\emph{arXiv preprint arXiv:2310.04378}}.
\newblock


\bibitem[\protect\citeauthoryear{Moser, Shanbhag, Raue, Frolov, Palacio, and Dengel}{Moser et~al\mbox{.}}{[n.d.]}]%
        {moser2401diffusion}
\bibfield{author}{\bibinfo{person}{BB Moser}, \bibinfo{person}{AS Shanbhag}, \bibinfo{person}{F Raue}, \bibinfo{person}{S Frolov}, \bibinfo{person}{S Palacio}, {and} \bibinfo{person}{A Dengel}.} \bibinfo{year}{[n.d.]}\natexlab{}.
\newblock \showarticletitle{Diffusion Models, Image Super-Resolution And Everything: A Survey. arXiv 2024}.
\newblock \bibinfo{journal}{\emph{arXiv preprint arXiv:2401.00736}}.
\newblock


\bibitem[\protect\citeauthoryear{Rombach, Blattmann, Lorenz, Esser, and Ommer}{Rombach et~al\mbox{.}}{2022}]%
        {rombach2022high}
\bibfield{author}{\bibinfo{person}{Robin Rombach}, \bibinfo{person}{Andreas Blattmann}, \bibinfo{person}{Dominik Lorenz}, \bibinfo{person}{Patrick Esser}, {and} \bibinfo{person}{Bj{\"o}rn Ommer}.} \bibinfo{year}{2022}\natexlab{}.
\newblock \showarticletitle{High-resolution image synthesis with latent diffusion models}. In \bibinfo{booktitle}{\emph{Proceedings of the IEEE/CVF conference on computer vision and pattern recognition}}. \bibinfo{pages}{10684--10695}.
\newblock


\bibitem[\protect\citeauthoryear{Wang, Yue, Zhou, Chan, and Loy}{Wang et~al\mbox{.}}{2024b}]%
        {wang2024exploiting}
\bibfield{author}{\bibinfo{person}{Jianyi Wang}, \bibinfo{person}{Zongsheng Yue}, \bibinfo{person}{Shangchen Zhou}, \bibinfo{person}{Kelvin~CK Chan}, {and} \bibinfo{person}{Chen~Change Loy}.} \bibinfo{year}{2024}\natexlab{b}.
\newblock \showarticletitle{Exploiting diffusion prior for real-world image super-resolution}.
\newblock \bibinfo{journal}{\emph{International Journal of Computer Vision}}, \bibinfo{pages}{1--21}.
\newblock


\bibitem[\protect\citeauthoryear{Wang, Xie, Dong, and Shan}{Wang et~al\mbox{.}}{2021}]%
        {realesrgan}
\bibfield{author}{\bibinfo{person}{Xintao Wang}, \bibinfo{person}{Liangbin Xie}, \bibinfo{person}{Chao Dong}, {and} \bibinfo{person}{Ying Shan}.} \bibinfo{year}{2021}\natexlab{}.
\newblock \showarticletitle{Real-ESRGAN: Training Real-World Blind Super-Resolution with Pure Synthetic Data}.
\newblock
\showeprint[arxiv]{2107.10833}~[eess.IV]
\urldef\tempurl%
\url{https://arxiv.org/abs/2107.10833}
\showURL{%
\tempurl}


\bibitem[\protect\citeauthoryear{Wang, Yang, Chen, Wang, Guo, Chau, Liu, Qiao, Kot, and Wen}{Wang et~al\mbox{.}}{2024a}]%
        {wang2024sinsr}
\bibfield{author}{\bibinfo{person}{Yufei Wang}, \bibinfo{person}{Wenhan Yang}, \bibinfo{person}{Xinyuan Chen}, \bibinfo{person}{Yaohui Wang}, \bibinfo{person}{Lanqing Guo}, \bibinfo{person}{Lap-Pui Chau}, \bibinfo{person}{Ziwei Liu}, \bibinfo{person}{Yu Qiao}, \bibinfo{person}{Alex~C Kot}, {and} \bibinfo{person}{Bihan Wen}.} \bibinfo{year}{2024}\natexlab{a}.
\newblock \showarticletitle{SinSR: diffusion-based image super-resolution in a single step}. In \bibinfo{booktitle}{\emph{Proceedings of the IEEE/CVF Conference on Computer Vision and Pattern Recognition}}. \bibinfo{pages}{25796--25805}.
\newblock


\bibitem[\protect\citeauthoryear{Xing, Dai, Hu, Wu, and Jiang}{Xing et~al\mbox{.}}{2024}]%
        {xing2024simda}
\bibfield{author}{\bibinfo{person}{Zhen Xing}, \bibinfo{person}{Qi Dai}, \bibinfo{person}{Han Hu}, \bibinfo{person}{Zuxuan Wu}, {and} \bibinfo{person}{Yu-Gang Jiang}.} \bibinfo{year}{2024}\natexlab{}.
\newblock \showarticletitle{Simda: Simple diffusion adapter for efficient video generation}. In \bibinfo{booktitle}{\emph{Proceedings of the IEEE/CVF Conference on Computer Vision and Pattern Recognition}}. \bibinfo{pages}{7827--7839}.
\newblock


\bibitem[\protect\citeauthoryear{Yue, Wang, and Loy}{Yue et~al\mbox{.}}{2024}]%
        {yue2024resshift}
\bibfield{author}{\bibinfo{person}{Zongsheng Yue}, \bibinfo{person}{Jianyi Wang}, {and} \bibinfo{person}{Chen~Change Loy}.} \bibinfo{year}{2024}\natexlab{}.
\newblock \showarticletitle{Resshift: Efficient diffusion model for image super-resolution by residual shifting}.
\newblock \bibinfo{journal}{\emph{Advances in Neural Information Processing Systems}}  \bibinfo{volume}{36} (\bibinfo{year}{2024}).
\newblock


\bibitem[\protect\citeauthoryear{Zhou, Yang, Wang, Luo, and Loy}{Zhou et~al\mbox{.}}{2024}]%
        {zhou2024upscale}
\bibfield{author}{\bibinfo{person}{Shangchen Zhou}, \bibinfo{person}{Peiqing Yang}, \bibinfo{person}{Jianyi Wang}, \bibinfo{person}{Yihang Luo}, {and} \bibinfo{person}{Chen~Change Loy}.} \bibinfo{year}{2024}\natexlab{}.
\newblock \showarticletitle{Upscale-A-Video: Temporal-Consistent Diffusion Model for Real-World Video Super-Resolution}. In \bibinfo{booktitle}{\emph{Proceedings of the IEEE/CVF Conference on Computer Vision and Pattern Recognition}}. \bibinfo{pages}{2535--2545}.
\newblock


\bibitem[\protect\citeauthoryear{Zhou and Wang}{Zhou and Wang}{2024}]%
        {zhou2024spatio}
\bibfield{author}{\bibinfo{person}{Tong Zhou} {and} \bibinfo{person}{Shuqiang Wang}.} \bibinfo{year}{2024}\natexlab{}.
\newblock \showarticletitle{Spatio-Temporal Adaptive Diffusion Models for EEG Super-Resolution in Epilepsy Diagnosis}.
\newblock \bibinfo{journal}{\emph{arXiv preprint arXiv:2407.03089}}.
\newblock


\end{thebibliography}

\end{document}